\documentclass[aps,prl,reprint,superscriptaddress,graphicx]{revtex4-1}

\usepackage{hyperref}
\usepackage{graphicx}
\usepackage{textcomp}
\usepackage{amsmath}
\usepackage{bm}
\usepackage{amssymb}
\usepackage{ulem}

\usepackage{color}

\newcommand{\tr}[1]{{\textrm{#1}}}

\begin{document}

\title{Optical second harmonic generation in encapsulated single-layer InSe}

\author{Nadine~Leisgang}
\email{nadine.leisgang@unibas.ch}
\affiliation{Department of Physics, University of Basel, Klingelbergstrasse 82, CH-4056 Basel, Switzerland}

\author{Jonas~G.~Roch}
\affiliation{Department of Physics, University of Basel, Klingelbergstrasse 82, CH-4056 Basel, Switzerland}

\author{Guillaume~Froehlicher}
\affiliation{Department of Physics, University of Basel, Klingelbergstrasse 82, CH-4056 Basel, Switzerland}

\author{Matthew~Hamer}
\affiliation{School of Physics, University of Manchester, Oxford Road, Manchester, M13 9PL, U.K.}
\affiliation{National Graphene Institute, University of Manchester, Oxford Road, Manchester, M13 9PL, U.K.}

\author{Daniel~Terry}
\affiliation{School of Physics, University of Manchester, Oxford Road, Manchester, M13 9PL, U.K.}
\affiliation{National Graphene Institute, University of Manchester, Oxford Road, Manchester, M13 9PL, U.K.}

\author{Roman~Gorbachev}
\affiliation{School of Physics, University of Manchester, Oxford Road, Manchester, M13 9PL, U.K.}
\affiliation{National Graphene Institute, University of Manchester, Oxford Road, Manchester, M13 9PL, U.K.}

\author{Richard~J.~Warburton}
\affiliation{Department of Physics, University of Basel, Klingelbergstrasse 82, CH-4056 Basel, Switzerland}


\begin{abstract}
We report the observation of optical second harmonic generation (SHG) in single-layer indium selenide (InSe). We measure a second harmonic signal of $>~10^3$~cts/s under nonresonant excitation using a home-built confocal microscope and a standard pulsed pico-second laser. We demonstrate that polarization-resolved SHG serves as a fast, non-invasive tool to determine the crystal axes in single-layer InSe and to relate the sharp edges of the flake to the armchair and zigzag edges of the crystal structure. Our experiment determines these angles to an accuracy better than $\pm$~0.2$^{\circ}$. Treating the two-dimensional material as a nonlinear polarizable sheet, we determine a second-order nonlinear sheet polarizability $| \chi_\tr{sheet}^{(2)}|=(17.9~\pm~11.0)\times 10^{-20}~\tr{m}^2~\tr{V}^{-1}$ for single-layer InSe, corresponding to an effective nonlinear susceptibility value of $| \chi_\tr{eff}^{(2)}| \approx (223~\pm~138)\times 10^{-12}~\tr{m}~\tr{V}^{-1}$ accounting for the sheet thickness (d~$\approx$~0.8~nm). We demonstrate that the SHG technique can also be applied to encapsulated samples to probe their crystal orientations. The method is therefore suitable for creating high quality van der Waals heterostructures with control over the crystal directions.
\end{abstract}


\maketitle
 
Since the re-discovery of graphene, two-dimensional (2D) materials, such as atomic layers of transition metal dichalcogenides (TMDCs) (e.g.\ MoS$_2$, MoSe$_2$, WS$_2$ and WSe$_2$) and III-IV compounds (e.g.\ GaS, GaSe) have attracted great attention in materials research on account of their tunable electronic and optical properties \cite{Xia2014,Mak2016}. The ability to combine the 2D layers with hexagonal boron nitride (h-BN) and few-layer graphene offers an opportunity of creating high-performance, 2D opto-electronic devices \cite{Dean2010,Kretinin2014}.

The desire to stack different 2D materials with precise control over the twist angle \cite{Mishchenko2014,Kunstmann2018,Liu2014,Zande2014} creates a need for a fast, non-invasive tool to probe the underlying crystal symmetries and crystallographic orientations. Nonlinear optical techniques, such as second harmonic generation (SHG), provide insight into the properties of surfaces or interfaces \cite{Shen2003}, particularly among non-centrosymmetric materials. In general, under an incident electric field \textbf{E}($\omega$) with fundamental angular frequency $\omega$, the second-order nonlinear polarization is determined by a third-rank electric susceptibility tensor
\begin{equation}
\chi^{(2)}: \, \tr{P}_i(2\omega)=\epsilon_0 \chi_{ijk}^{(2)}(2 \omega;\omega,\omega)\tr{E}_j(\omega)\tr{E}_k(\omega)
\end{equation}
resulting in the creation of SHG. The second-order nonlinearities of a material can be probed by impinging intense linearly polarized light at angular frequency $\omega$ onto its surface and measuring the generated outgoing response at $2 \omega$ through an analyzer. By using different polarization combinations and/or by varying the orientation of the interface with respect to the incoming beam, different components of $\chi^{(2)}$ can be determined, giving information about the structural symmetry of the material as well as the strength of various nonlinear processes. SHG has been observed in single- and few-layer TMDCs, such as MoS$_2$ \cite{Kumar2013,Malard2013} and WS$_2$ \cite{Janisch2014}, and has been used to align the crystal axes of various TMDCs in order to create high-quality heterobilayers with strong interlayer exciton emission \cite{Rivera2015,Nayak2017}. However, TMDC flakes with an even number of layers exhibit inversion symmetry, resulting in a vanishing second-order nonlinearity ($\chi^{(2)}=0$) which prevents the observation of SHG for all layer thicknesses.

\begin{figure}[!t]
\begin{center}
\includegraphics[width=8.5cm]{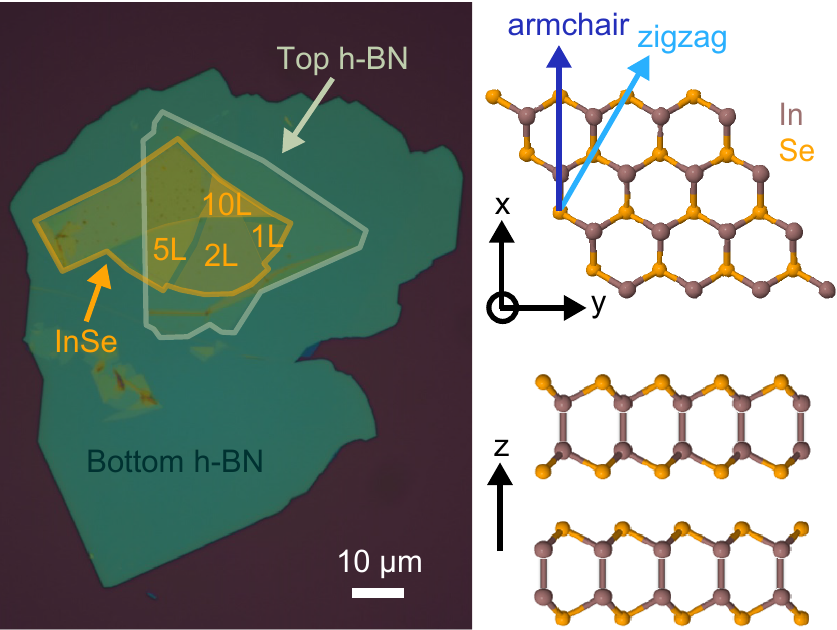}
\caption{Left:~Optical image of the InSe flake. The indicated layer thicknesses were determined by optical contrast and verified by atomic force microscopy and photoluminescence measurements. Second harmonic generation measurements were carried out on the parts of the InSe flake which are fully encapsulated in h-BN. The device consists of SiO$_2$ (290~nm)/h-BN (46~nm)/InSe/h-BN (8~nm). Right:~Schematic of the InSe crystal structure (top and side view). Purple and orange spheres correspond to indium (In) and selenium (Se) atoms, respectively.}
\label{Fig1}
\end{center}
\end{figure}

Among the large family of van der Waals (vdW) crystals, indium selenide (InSe) has emerged as a promising 2D semiconductor due to its highly tunable optical response in the near-infrared to the visible spectrum \cite{Mudd2016} and its high electron mobilities at room and liquid-helium temperatures allowing the quantum Hall effect to be observed \cite{Bandurin2016}. Similar to the TMDCs, single-layer InSe belongs to the non-centrosymmetric D$_{3 \tr{h}}$ ($\bar{6}$m2) point group with only one independent non-zero second-order nonlinear susceptibility tensor element. The susceptibility components satisfy $\chi_{xxx}^{(2)}=-\chi_{xyy}^{(2)}=-\chi_{yyx}^{(2)}=\chi_{yxy}^{(2)}$ with $x$ along the armchair direction (Fig.~\ref{Fig1} (right)) \cite{Shen2003}. In contrast, the specific stacking order in bulk and few-layer InSe breaks the mirror-plane symmetry characteristic of single-layer InSe, thus maintaining broken inversion symmetry for all layer thicknesses. This provides an opportunity to investigate nonlinearities in thin InSe films layer-by-layer. The dependence of polarization-resolved SHG on the crystallographic axes further opens up an optical means of characterizing the crystal structure and orientation of the thin InSe films. Indeed, observations of SHG in relatively thin InSe sheets (from 9 to 25~nm) with even and odd number of layers have been reported recently \cite{Deckoff2016}. However, in the single-layer limit and under non-resonant condition, it remains elusive whether SHG signal can still be detected. In this Letter, we present SHG from encapsulated single- and few-layer InSe with the aim of determining the crystal axes and estimating the effective second-order nonlinear susceptibility of the single-layer.

Thin unprotected InSe films have optical properties which deteriorate over time, interpreted as a gradual degradation of the crystal due to interaction with oxygen and water in the atmosphere \cite{Zamudio2015,Bandurin2016}. Thus, to protect the material from the interaction with the environment, we employed exfoliation and subsequent encapsulation of single- and few-layer InSe in an inert (argon) atmosphere. The resulting InSe structures are stable under ambient conditions even in the single-layer limit \cite{Cao2015}. Fig.~\ref{Fig1} (left) shows an optical image of the sample used to investigate the SHG response of thin InSe flakes. Layers of different thickness and single-layer steps ($\approx$~0.8~nm) have been identified by atomic force microscopy and photoluminescence measurements.

SHG spectroscopy was performed at room temperature using a home-built, confocal microscope setup. The optical pulse, centered at a wavelength of 810~nm, was obtained from a Ti:Sapphire laser with 76~MHz repetition rate. All the spectral components of the $\sim$~150~fs pulse were retained; dispersion in the optical fiber connecting the laser source to the microscope stretches the pulse to the pico-second domain at the sample (intensity FWHM 35.9~ps). The laser (average power 3.2~mW) was focused to a spot size of about 1.5~\textmu m on the sample by a microscope objective lens (40x, NA~$=$~0.65) at normal incidence and with a fixed linear polarization. The SHG signal was collected by the same objective and directed through a dichroic beamsplitter to a spectrometer equipped with a 300 grooves/mm grating and a nitrogen-cooled silicon charge-coupled device (CCD).

\begin{figure}[!b]
\begin{center}
\includegraphics[width=8.5cm]{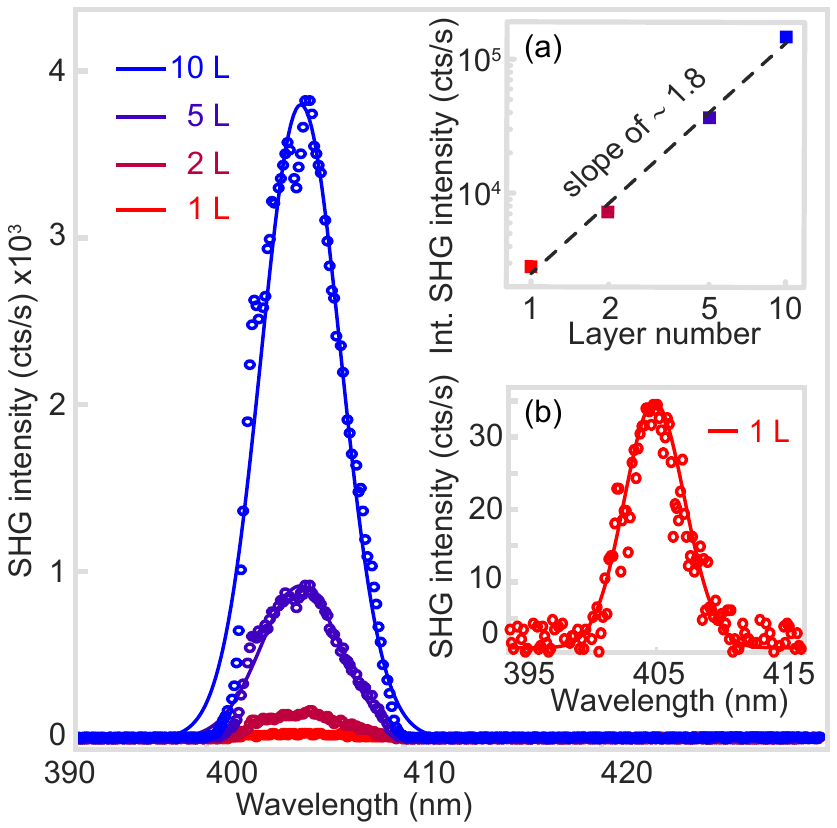}
\caption{SHG spectra of single- and few-layer InSe at room temperature. The inset~(a) (plotted on a double logarithmic scale) shows the quadratic increase (slope of $\sim$~1.8) of the integrated SHG with the number of layers. The error bars in (a) are smaller than the symbol size. The inset~(b) shows the SHG spectrum of encapsulated single-layer InSe.}
\label{Fig2}
\end{center}
\end{figure}

\begin{figure*}[!t]
\begin{center}
\includegraphics[width=17cm]{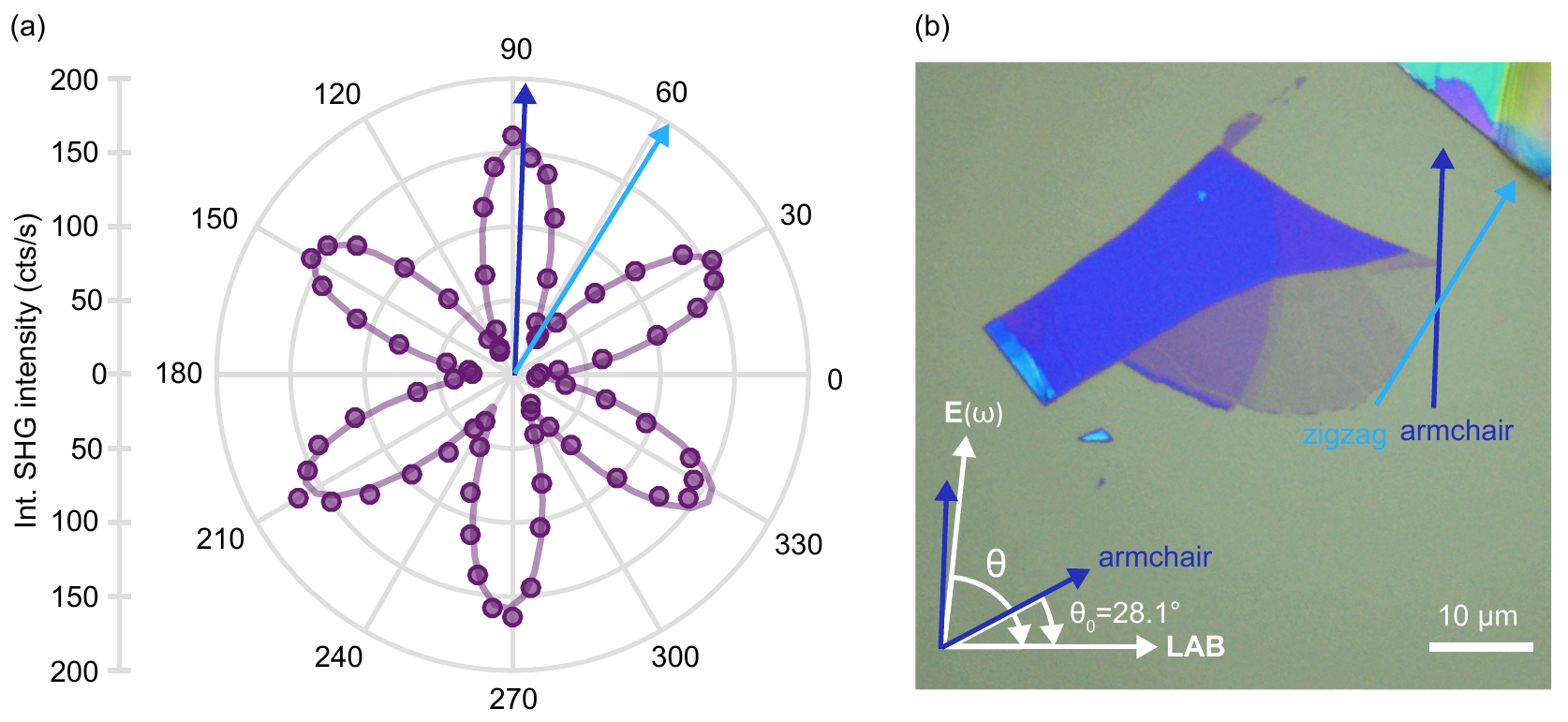}
\caption{SHG characterization of encapsulated InSe. (a) Polar plot of the ``parallel" SHG intensity I$^{\parallel}_\tr{SHG}(2\omega)$ of single-layer InSe as a function of rotation angle $\theta$. Fitting the angular dependence to I$^{\parallel}_\tr{SHG}(2\omega)\propto \cos^2 3(\theta-\theta_0)$ (solid purple line), the armchair direction (dark blue arrow) of the sample is determined as $\theta_0=28.1^{\circ}~\pm~0.2 ^{\circ}$. This armchair direction is 30$^{\circ}$ from the zigzag direction (light blue arrow). (b) Optical image of the InSe flakes before encapsulation. The armchair direction forms an angle $\theta_0$ with respect to the ``laboratory" axis (LAB). The indicated armchair (dark blue arrow) and zigzag (light blue arrow) edges were determined by polarization-resolved SHG performed on the encapsulated sample.}
\label{Fig3}
\end{center}
\end{figure*}

As shown in Fig.~\ref{Fig2}, an easy-to-measure SHG signal ($> 10^3$~cts/s) appeared at wavelength 405~nm when the laser beam (wavelength 810~nm) was focussed on encapsulated single- and few-layer InSe. Moreover, the SHG response could be observed for both even and odd number of layers. No observable SHG signal was measured on h-BN. The integrated SHG signal increases by more than two orders of magnitude as the thickness of the InSe flake increases from 1 to 10 layers (inset~(a) to Fig.~\ref{Fig2}). Specifically, the SHG signal depends quadratically on the layer thickness, i.e.\ $\tr{I}_\tr{SHG}(2\omega) \propto \tr{d}^2$, as shown in the inset~(a) in Fig.~\ref{Fig2} where the integrated SHG intensity I$_\tr{SHG}(2\omega)$ of the InSe flakes is plotted versus the layer number N (with d~$\approx$~N~$\times$~0.8~nm) on a logarithmic scale. Following \cite{Yariv1984}, the dependence of the SHG intensity on the flake thickness is given by
\begin{equation}
\tr{I}_\tr{SHG}(2\omega)\propto (\tr{l}_\tr{c} | \chi_\tr{eff}^{(2)}|)^2 \sin^2\left(\frac{\pi \tr{d}}{2\tr{l}_\tr{c}}\right)
\label{ThicknessDependence}
\end{equation}
where d is the flake thickness, $\tr{l}_\tr{c}=2\pi/\Delta k \approx 20$~\textmu m is the coherence length and $| \chi_\tr{eff}^{(2)}|$ is the effective nonlinear susceptibility determined by the sample geometry and the non-zero components of the nonlinear sheet susceptibility tensor $\chi_\tr{sheet}^{(2)}$ ($| \chi_\tr{eff}^{(2)}|=| \chi_\tr{sheet}^{(2)}|/$d). Thus, in the limit of atomically thin films (i.e. d $\ll$ l$_\tr{c}$), the SHG intensity I$_\tr{SHG}(2\omega)$ is expected to increase quadratically with the number of layers, in good agreement with our observations (inset~(a) to Fig.~\ref{Fig2}).

The relatively strong SHG signal facilitates an investigation of the lattice symmetry and crystallographic orientation of the thin flakes. The SHG intensity I$_\tr{SHG}(2\omega)$ is strongly dependent on the polarization angle $(\theta-\theta_0)$ between the laser polarization \textbf{E}($\omega$) and the armchair direction of the crystal defined in Fig.~\ref{Fig3}(b). For polarization-resolved SHG, the laser polarization was rotated about the z-axis by a half-wave plate to vary $\theta$, and the SHG signal I$^{\parallel}_\tr{SHG}(2\omega)$ (SHG polarization parallel to the excitation polarization) was collected using an analyzer located in front of the detector. The polar plot for single-layer InSe in Fig.~\ref{Fig3}(a) shows a strongly varying 6-fold symmetry of I$^{\parallel}_\tr{SHG}(2\omega)\propto \cos^2 3(\theta-\theta_0)$. This directly reveals the underlying symmetry and orientation of the single-layer InSe flake. The initial orientation of the sample with respect to the armchair direction of the crystal was determined to be $\theta_0=28.1^{\circ}~\pm~0.2 ^{\circ}$. The sharp edges along which the single-layer InSe flake cleaved during exfoliation could therefore be clearly assigned to the armchair and zigzag crystal axes, respectively (Fig.~\ref{Fig3}(b)).

To quantify the nonlinear response of single-layer InSe, we follow the formalism of \cite{Shen1989} where SHG from an ultrathin layer is treated as radiation from a nonlinear, polarizable sheet embedded in a dielectric medium, with boundary effects taken into account by Fresnel transmission coefficients. The second-order susceptibility $| \chi_\tr{sheet}^{(2)}|$ of InSe can be extracted from measurements of the intensity of the SHG with respect to the driving intensity via
\begin{equation}
\tr{I}_\tr{SHG}(2\omega)=\frac{\omega^2}{2c^3\epsilon_0}\tr{t}_{\tr{in}}^4(\omega) \tr{t}_{\tr{out}}^2(2\omega) | \chi_\tr{sheet}^{(2)}|^2 \tr{I}^2(\omega)
\label{NonlinearSusceptibility}
\end{equation}
where t$_{\tr{in}}(\omega)$ and t$_{\tr{out}}(2\omega)$ account for the local field correction factors due to the dielectric environment determined by transfer matrix methods (``Essential Macleod''). Relating the laser intensity I$(\omega)$ and the SHG intensity I$_\tr{SHG}(2\omega)$ in Eq.~(\ref{NonlinearSusceptibility}) to the experimentally measured time-averaged power values, we obtain $| \chi_\tr{sheet}^{(2)}|=(17.9~\pm~11.0)\times 10^{-20}~\tr{m}^2~\tr{V}^{-1}$ for single-layer InSe. To compare to other nonlinear optical bulk materials, we estimate an effective bulk-like nonlinear susceptibility $| \chi_\tr{eff}^{(2)}|=(223~\pm~138)\times 10^{-12}~\tr{m}~\tr{V}^{-1}$ by including the thickness of the 2D material (d~$\approx$~0.8~nm). This value is similar to the strong second-order optical susceptibility measured for single- and few-layer GaSe \cite{Tang2016}.

In conclusion, we report an observation of SHG in single-layer InSe under nonresonant excitation, yielding a nonlinear sheet susceptibility with an estimated value of $|\chi_\tr{eff}^{(2)}| \approx 223~\tr{pm}~\tr{V}^{-1}$. Quantitative characterization of the nonlinear response of single- and few-layer InSe reveals the expected quadratic dependence of the SHG signal on the number of layers. The crystalline symmetry was probed by polarization-resolved SHG where the ``petal" direction with maximum signal is parallel to the in-plane In-Se or Se-In (armchair) direction. This allowed for a fast and precise ($\pm~0.2 ^{\circ}$) assignment of the sharp edges of the InSe flake to its crystal axes, demonstrating that SHG serves as useful tool for the determination of the orientation of the material's crystallographic axes. The ability to distinguish different crystallographic axes, even in encapsulated samples, can be exploited to improve the quality of van der Waals heterostructures by stacking various 2D materials with precise twist-angle control.

\begin{acknowledgements}
We thank Immo~S\"ollner, Jan-Philipp~Jahn and Daniel~Najer for technical assistance and experiments. The work in Basel was financially supported by SNF [Project Number 200020\_175748], Swiss Nanoscience Institute (SNI), QCQT PhD School and NCCR QSIT. The work in Manchester was financially supported by the Engineering and Physical Sciences Research Council [Grant Number EP/M012700/1] (EPSRC). R.V.G. acknowledges financial support from the Royal Society Fellowship Scheme. M.H. and D.T. acknowledge support from the EPSRC NowNano Doctoral Training Centre.
\end{acknowledgements}


\begin{thebibliography}{23}
\makeatletter
\providecommand \@ifxundefined [1]{%
 \@ifx{#1\undefined}
}%
\providecommand \@ifnum [1]{%
 \ifnum #1\expandafter \@firstoftwo
 \else \expandafter \@secondoftwo
 \fi
}%
\providecommand \@ifx [1]{%
 \ifx #1\expandafter \@firstoftwo
 \else \expandafter \@secondoftwo
 \fi
}%
\providecommand \natexlab [1]{#1}%
\providecommand \enquote  [1]{``#1''}%
\providecommand \bibnamefont  [1]{#1}%
\providecommand \bibfnamefont [1]{#1}%
\providecommand \citenamefont [1]{#1}%
\providecommand \href@noop [0]{\@secondoftwo}%
\providecommand \href [0]{\begingroup \@sanitize@url \@href}%
\providecommand \@href[1]{\@@startlink{#1}\@@href}%
\providecommand \@@href[1]{\endgroup#1\@@endlink}%
\providecommand \@sanitize@url [0]{\catcode `\\12\catcode `\$12\catcode
  `\&12\catcode `\#12\catcode `\^12\catcode `\_12\catcode `\%12\relax}%
\providecommand \@@startlink[1]{}%
\providecommand \@@endlink[0]{}%
\providecommand \url  [0]{\begingroup\@sanitize@url \@url }%
\providecommand \@url [1]{\endgroup\@href {#1}{\urlprefix }}%
\providecommand \urlprefix  [0]{URL }%
\providecommand \Eprint [0]{\href }%
\providecommand \doibase [0]{http://dx.doi.org/}%
\providecommand \selectlanguage [0]{\@gobble}%
\providecommand \bibinfo  [0]{\@secondoftwo}%
\providecommand \bibfield  [0]{\@secondoftwo}%
\providecommand \translation [1]{[#1]}%
\providecommand \BibitemOpen [0]{}%
\providecommand \bibitemStop [0]{}%
\providecommand \bibitemNoStop [0]{.\EOS\space}%
\providecommand \EOS [0]{\spacefactor3000\relax}%
\providecommand \BibitemShut  [1]{\csname bibitem#1\endcsname}%
\let\auto@bib@innerbib\@empty
\bibitem [{\citenamefont {Xia}\ \emph {et~al.}(2014)\citenamefont {Xia},
  \citenamefont {Wang}, \citenamefont {Xiao}, \citenamefont {Dubey},\ and\
  \citenamefont {Ramasubramaniam}}]{Xia2014}%
  \BibitemOpen
  \bibfield  {author} {\bibinfo {author} {\bibfnamefont {F.}~\bibnamefont
  {Xia}}, \bibinfo {author} {\bibfnamefont {H.}~\bibnamefont {Wang}}, \bibinfo
  {author} {\bibfnamefont {D.}~\bibnamefont {Xiao}}, \bibinfo {author}
  {\bibfnamefont {M.}~\bibnamefont {Dubey}}, \ and\ \bibinfo {author}
  {\bibfnamefont {A.}~\bibnamefont {Ramasubramaniam}},\ }\href
  {http://dx.doi.org/10.1038/nphoton.2014.271} {\bibfield  {journal} {\bibinfo
  {journal} {Nat. Photonics}\ }\textbf {\bibinfo {volume} {8}},\ \bibinfo
  {pages} {899} (\bibinfo {year} {2014})}\BibitemShut {NoStop}%
\bibitem [{\citenamefont {Mak}\ and\ \citenamefont {Shan}(2016)}]{Mak2016}%
  \BibitemOpen
  \bibfield  {author} {\bibinfo {author} {\bibfnamefont {K.~F.}\ \bibnamefont
  {Mak}}\ and\ \bibinfo {author} {\bibfnamefont {J.}~\bibnamefont {Shan}},\
  }\href {http://dx.doi.org/10.1038/nphoton.2015.282} {\bibfield  {journal}
  {\bibinfo  {journal} {Nat. Photonics}\ }\textbf {\bibinfo {volume} {10}},\
  \bibinfo {pages} {216} (\bibinfo {year} {2016})}\BibitemShut {NoStop}%
\bibitem [{\citenamefont {Dean}\ \emph {et~al.}(2010)\citenamefont {Dean},
  \citenamefont {Young}, \citenamefont {Meric}, \citenamefont {Lee},
  \citenamefont {Wang}, \citenamefont {Sorgenfrei}, \citenamefont {Watanabe},
  \citenamefont {Taniguchi}, \citenamefont {Kim}, \citenamefont {Shepard},\
  and\ \citenamefont {Hone}}]{Dean2010}%
  \BibitemOpen
  \bibfield  {author} {\bibinfo {author} {\bibfnamefont {C.~R.}\ \bibnamefont
  {Dean}}, \bibinfo {author} {\bibfnamefont {A.~F.}\ \bibnamefont {Young}},
  \bibinfo {author} {\bibfnamefont {I.}~\bibnamefont {Meric}}, \bibinfo
  {author} {\bibfnamefont {C.}~\bibnamefont {Lee}}, \bibinfo {author}
  {\bibfnamefont {L.}~\bibnamefont {Wang}}, \bibinfo {author} {\bibfnamefont
  {S.}~\bibnamefont {Sorgenfrei}}, \bibinfo {author} {\bibfnamefont
  {K.}~\bibnamefont {Watanabe}}, \bibinfo {author} {\bibfnamefont
  {T.}~\bibnamefont {Taniguchi}}, \bibinfo {author} {\bibfnamefont
  {P.}~\bibnamefont {Kim}}, \bibinfo {author} {\bibfnamefont {K.~L.}\
  \bibnamefont {Shepard}}, \ and\ \bibinfo {author} {\bibfnamefont
  {J.}~\bibnamefont {Hone}},\ }\href {http://dx.doi.org/10.1038/nnano.2010.172}
  {\bibfield  {journal} {\bibinfo  {journal} {Nat. Nanotechnol.}\ }\textbf
  {\bibinfo {volume} {5}},\ \bibinfo {pages} {722} (\bibinfo {year}
  {2010})}\BibitemShut {NoStop}%
\bibitem [{\citenamefont {Kretinin}\ \emph {et~al.}(2014)\citenamefont
  {Kretinin}, \citenamefont {Cao}, \citenamefont {Tu}, \citenamefont {Yu},
  \citenamefont {Jalil}, \citenamefont {Novoselov}, \citenamefont {Haigh},
  \citenamefont {Gholinia}, \citenamefont {Mishchenko}, \citenamefont {Lozada},
  \citenamefont {Georgiou}, \citenamefont {Woods}, \citenamefont {Withers},
  \citenamefont {Blake}, \citenamefont {Eda}, \citenamefont {Wirsig},
  \citenamefont {Hucho}, \citenamefont {Watanabe}, \citenamefont {Taniguchi},
  \citenamefont {Geim},\ and\ \citenamefont {Gorbachev}}]{Kretinin2014}%
  \BibitemOpen
  \bibfield  {author} {\bibinfo {author} {\bibfnamefont {A.~V.}\ \bibnamefont
  {Kretinin}}, \bibinfo {author} {\bibfnamefont {Y.}~\bibnamefont {Cao}},
  \bibinfo {author} {\bibfnamefont {J.~S.}\ \bibnamefont {Tu}}, \bibinfo
  {author} {\bibfnamefont {G.~L.}\ \bibnamefont {Yu}}, \bibinfo {author}
  {\bibfnamefont {R.}~\bibnamefont {Jalil}}, \bibinfo {author} {\bibfnamefont
  {K.~S.}\ \bibnamefont {Novoselov}}, \bibinfo {author} {\bibfnamefont {S.~J.}\
  \bibnamefont {Haigh}}, \bibinfo {author} {\bibfnamefont {A.}~\bibnamefont
  {Gholinia}}, \bibinfo {author} {\bibfnamefont {A.}~\bibnamefont
  {Mishchenko}}, \bibinfo {author} {\bibfnamefont {M.}~\bibnamefont {Lozada}},
  \bibinfo {author} {\bibfnamefont {T.}~\bibnamefont {Georgiou}}, \bibinfo
  {author} {\bibfnamefont {C.~R.}\ \bibnamefont {Woods}}, \bibinfo {author}
  {\bibfnamefont {F.}~\bibnamefont {Withers}}, \bibinfo {author} {\bibfnamefont
  {P.}~\bibnamefont {Blake}}, \bibinfo {author} {\bibfnamefont
  {G.}~\bibnamefont {Eda}}, \bibinfo {author} {\bibfnamefont {A.}~\bibnamefont
  {Wirsig}}, \bibinfo {author} {\bibfnamefont {C.}~\bibnamefont {Hucho}},
  \bibinfo {author} {\bibfnamefont {K.}~\bibnamefont {Watanabe}}, \bibinfo
  {author} {\bibfnamefont {T.}~\bibnamefont {Taniguchi}}, \bibinfo {author}
  {\bibfnamefont {A.~K.}\ \bibnamefont {Geim}}, \ and\ \bibinfo {author}
  {\bibfnamefont {R.~V.}\ \bibnamefont {Gorbachev}},\ }\href {\doibase
  10.1021/nl5006542} {\bibfield  {journal} {\bibinfo  {journal} {Nano Lett.}\
  }\textbf {\bibinfo {volume} {14}},\ \bibinfo {pages} {3270} (\bibinfo {year}
  {2014})}\BibitemShut {NoStop}%
\bibitem [{\citenamefont {Mishchenko}\ \emph {et~al.}(2014)\citenamefont
  {Mishchenko}, \citenamefont {Tu}, \citenamefont {Cao}, \citenamefont
  {Gorbachev}, \citenamefont {Wallbank}, \citenamefont {Greenaway},
  \citenamefont {Morozov}, \citenamefont {Morozov}, \citenamefont {Zhu},
  \citenamefont {Wong}, \citenamefont {Withers}, \citenamefont {Woods},
  \citenamefont {Kim}, \citenamefont {Watanabe}, \citenamefont {Taniguchi},
  \citenamefont {Vdovin}, \citenamefont {Makarovsky}, \citenamefont {Fromhold},
  \citenamefont {Fal'ko}, \citenamefont {Geim}, \citenamefont {Eaves},\ and\
  \citenamefont {Novoselov}}]{Mishchenko2014}%
  \BibitemOpen
  \bibfield  {author} {\bibinfo {author} {\bibfnamefont {A.}~\bibnamefont
  {Mishchenko}}, \bibinfo {author} {\bibfnamefont {J.~S.}\ \bibnamefont {Tu}},
  \bibinfo {author} {\bibfnamefont {Y.}~\bibnamefont {Cao}}, \bibinfo {author}
  {\bibfnamefont {R.~V.}\ \bibnamefont {Gorbachev}}, \bibinfo {author}
  {\bibfnamefont {J.~R.}\ \bibnamefont {Wallbank}}, \bibinfo {author}
  {\bibfnamefont {M.~T.}\ \bibnamefont {Greenaway}}, \bibinfo {author}
  {\bibfnamefont {V.~E.}\ \bibnamefont {Morozov}}, \bibinfo {author}
  {\bibfnamefont {S.~V.}\ \bibnamefont {Morozov}}, \bibinfo {author}
  {\bibfnamefont {M.~J.}\ \bibnamefont {Zhu}}, \bibinfo {author} {\bibfnamefont
  {S.~L.}\ \bibnamefont {Wong}}, \bibinfo {author} {\bibfnamefont
  {F.}~\bibnamefont {Withers}}, \bibinfo {author} {\bibfnamefont {C.~R.}\
  \bibnamefont {Woods}}, \bibinfo {author} {\bibfnamefont {Y.-J.}\ \bibnamefont
  {Kim}}, \bibinfo {author} {\bibfnamefont {K.}~\bibnamefont {Watanabe}},
  \bibinfo {author} {\bibfnamefont {T.}~\bibnamefont {Taniguchi}}, \bibinfo
  {author} {\bibfnamefont {E.~E.}\ \bibnamefont {Vdovin}}, \bibinfo {author}
  {\bibfnamefont {O.}~\bibnamefont {Makarovsky}}, \bibinfo {author}
  {\bibfnamefont {T.~M.}\ \bibnamefont {Fromhold}}, \bibinfo {author}
  {\bibfnamefont {V.~I.}\ \bibnamefont {Fal'ko}}, \bibinfo {author}
  {\bibfnamefont {A.~K.}\ \bibnamefont {Geim}}, \bibinfo {author}
  {\bibfnamefont {L.}~\bibnamefont {Eaves}}, \ and\ \bibinfo {author}
  {\bibfnamefont {K.~S.}\ \bibnamefont {Novoselov}},\ }\href
  {http://dx.doi.org/10.1038/nnano.2014.187} {\bibfield  {journal} {\bibinfo
  {journal} {Nat. Nanotechnol.}\ }\textbf {\bibinfo {volume} {9}},\
  \bibinfo {pages} {808} (\bibinfo {year} {2014})}\BibitemShut {NoStop}%
\bibitem [{\citenamefont {Kunstmann}\ \emph {et~al.}(2018)\citenamefont
  {Kunstmann}, \citenamefont {Mooshammer}, \citenamefont {Nagler},
  \citenamefont {Chaves}, \citenamefont {Stein}, \citenamefont {Paradiso},
  \citenamefont {Plechinger}, \citenamefont {Strunk}, \citenamefont
  {Sch{\"u}ller}, \citenamefont {Seifert}, \citenamefont {Reichman},\ and\
  \citenamefont {Korn}}]{Kunstmann2018}%
  \BibitemOpen
  \bibfield  {author} {\bibinfo {author} {\bibfnamefont {J.}~\bibnamefont
  {Kunstmann}}, \bibinfo {author} {\bibfnamefont {F.}~\bibnamefont
  {Mooshammer}}, \bibinfo {author} {\bibfnamefont {P.}~\bibnamefont {Nagler}},
  \bibinfo {author} {\bibfnamefont {A.}~\bibnamefont {Chaves}}, \bibinfo
  {author} {\bibfnamefont {F.}~\bibnamefont {Stein}}, \bibinfo {author}
  {\bibfnamefont {N.}~\bibnamefont {Paradiso}}, \bibinfo {author}
  {\bibfnamefont {G.}~\bibnamefont {Plechinger}}, \bibinfo {author}
  {\bibfnamefont {C.}~\bibnamefont {Strunk}}, \bibinfo {author} {\bibfnamefont
  {C.}~\bibnamefont {Sch{\"u}ller}}, \bibinfo {author} {\bibfnamefont
  {G.}~\bibnamefont {Seifert}}, \bibinfo {author} {\bibfnamefont {D.~R.}\
  \bibnamefont {Reichman}}, \ and\ \bibinfo {author} {\bibfnamefont
  {T.}~\bibnamefont {Korn}},\ }\href
  {https://doi.org/10.1038/s41567-018-0123-y} {\bibfield  {journal} {\bibinfo
  {journal} {Nat. Phys.}\ }\textbf {\bibinfo {volume} {14}},\ \bibinfo
  {pages} {801} (\bibinfo {year} {2018})}\BibitemShut {NoStop}%
\bibitem [{\citenamefont {Liu}\ \emph {et~al.}(2014)\citenamefont {Liu},
  \citenamefont {Zhang}, \citenamefont {Cao}, \citenamefont {Jin},
  \citenamefont {Qiu}, \citenamefont {Zhou}, \citenamefont {Zettl},
  \citenamefont {Yang}, \citenamefont {Louie},\ and\ \citenamefont
  {Wang}}]{Liu2014}%
  \BibitemOpen
  \bibfield  {author} {\bibinfo {author} {\bibfnamefont {K.}~\bibnamefont
  {Liu}}, \bibinfo {author} {\bibfnamefont {L.}~\bibnamefont {Zhang}}, \bibinfo
  {author} {\bibfnamefont {T.}~\bibnamefont {Cao}}, \bibinfo {author}
  {\bibfnamefont {C.}~\bibnamefont {Jin}}, \bibinfo {author} {\bibfnamefont
  {D.}~\bibnamefont {Qiu}}, \bibinfo {author} {\bibfnamefont {Q.}~\bibnamefont
  {Zhou}}, \bibinfo {author} {\bibfnamefont {A.}~\bibnamefont {Zettl}},
  \bibinfo {author} {\bibfnamefont {P.}~\bibnamefont {Yang}}, \bibinfo {author}
  {\bibfnamefont {S.~G.}\ \bibnamefont {Louie}}, \ and\ \bibinfo {author}
  {\bibfnamefont {F.}~\bibnamefont {Wang}},\ }\href
  {http://dx.doi.org/10.1038/ncomms5966} {\bibfield  {journal} {\bibinfo
  {journal} {Nat. Commun.}\ }\textbf {\bibinfo {volume} {5}},\
  \bibinfo {pages} {4966} (\bibinfo {year} {2014})}\BibitemShut {NoStop}%
\bibitem [{\citenamefont {van~der Zande}\ \emph {et~al.}(2014)\citenamefont
  {van~der Zande}, \citenamefont {Kunstmann}, \citenamefont {Chernikov},
  \citenamefont {Chenet}, \citenamefont {You}, \citenamefont {Zhang},
  \citenamefont {Huang}, \citenamefont {Berkelbach}, \citenamefont {Wang},
  \citenamefont {Zhang}, \citenamefont {Hybertsen}, \citenamefont {Muller},
  \citenamefont {Reichman}, \citenamefont {Heinz},\ and\ \citenamefont
  {Hone}}]{Zande2014}%
  \BibitemOpen
  \bibfield  {author} {\bibinfo {author} {\bibfnamefont {A.~M.}\ \bibnamefont
  {van~der Zande}}, \bibinfo {author} {\bibfnamefont {J.}~\bibnamefont
  {Kunstmann}}, \bibinfo {author} {\bibfnamefont {A.}~\bibnamefont
  {Chernikov}}, \bibinfo {author} {\bibfnamefont {D.~A.}\ \bibnamefont
  {Chenet}}, \bibinfo {author} {\bibfnamefont {Y.}~\bibnamefont {You}},
  \bibinfo {author} {\bibfnamefont {X.}~\bibnamefont {Zhang}}, \bibinfo
  {author} {\bibfnamefont {P.~Y.}\ \bibnamefont {Huang}}, \bibinfo {author}
  {\bibfnamefont {T.~C.}\ \bibnamefont {Berkelbach}}, \bibinfo {author}
  {\bibfnamefont {L.}~\bibnamefont {Wang}}, \bibinfo {author} {\bibfnamefont
  {F.}~\bibnamefont {Zhang}}, \bibinfo {author} {\bibfnamefont {M.~S.}\
  \bibnamefont {Hybertsen}}, \bibinfo {author} {\bibfnamefont {D.~A.}\
  \bibnamefont {Muller}}, \bibinfo {author} {\bibfnamefont {D.~R.}\
  \bibnamefont {Reichman}}, \bibinfo {author} {\bibfnamefont {T.~F.}\
  \bibnamefont {Heinz}}, \ and\ \bibinfo {author} {\bibfnamefont {J.~C.}\
  \bibnamefont {Hone}},\ }\href {https://doi.org/10.1021/nl501077m} {\bibfield
  {journal} {\bibinfo  {journal} {Nano Lett.}\ }\textbf {\bibinfo {volume}
  {14}},\ \bibinfo {pages} {3869} (\bibinfo {year} {2014})}\BibitemShut
  {NoStop}%
\bibitem [{\citenamefont {Shen}(2003)}]{Shen2003}%
  \BibitemOpen
  \bibfield  {author} {\bibinfo {author} {\bibfnamefont {Y.~R.}\ \bibnamefont
  {Shen}},\ }\enquote {\bibinfo {title} {The principles of nonlinear optics},}\
  \ (\bibinfo  {publisher} {Wiley},\ \bibinfo {address} {New York},\ \bibinfo
  {year} {2003})\BibitemShut {NoStop}%
\bibitem [{\citenamefont {Kumar}\ \emph {et~al.}(2013)\citenamefont {Kumar},
  \citenamefont {Najmaei}, \citenamefont {Cui}, \citenamefont {Ceballos},
  \citenamefont {Ajayan}, \citenamefont {Lou},\ and\ \citenamefont
  {Zhao}}]{Kumar2013}%
  \BibitemOpen
  \bibfield  {author} {\bibinfo {author} {\bibfnamefont {N.}~\bibnamefont
  {Kumar}}, \bibinfo {author} {\bibfnamefont {S.}~\bibnamefont {Najmaei}},
  \bibinfo {author} {\bibfnamefont {Q.}~\bibnamefont {Cui}}, \bibinfo {author}
  {\bibfnamefont {F.}~\bibnamefont {Ceballos}}, \bibinfo {author}
  {\bibfnamefont {P.~M.}\ \bibnamefont {Ajayan}}, \bibinfo {author}
  {\bibfnamefont {J.}~\bibnamefont {Lou}}, \ and\ \bibinfo {author}
  {\bibfnamefont {H.}~\bibnamefont {Zhao}},\ }\href {\doibase
  10.1103/PhysRevB.87.161403} {\bibfield  {journal} {\bibinfo  {journal} {Phys.
  Rev. B}\ }\textbf {\bibinfo {volume} {87}},\ \bibinfo {pages} {161403}
  (\bibinfo {year} {2013})}\BibitemShut {NoStop}%
\bibitem [{\citenamefont {Malard}\ \emph {et~al.}(2013)\citenamefont {Malard},
  \citenamefont {Alencar}, \citenamefont {Barboza}, \citenamefont {Mak},\ and\
  \citenamefont {de~Paula}}]{Malard2013}%
  \BibitemOpen
  \bibfield  {author} {\bibinfo {author} {\bibfnamefont {L.~M.}\ \bibnamefont
  {Malard}}, \bibinfo {author} {\bibfnamefont {T.~V.}\ \bibnamefont {Alencar}},
  \bibinfo {author} {\bibfnamefont {A.~P.~M.}\ \bibnamefont {Barboza}},
  \bibinfo {author} {\bibfnamefont {K.~F.}\ \bibnamefont {Mak}}, \ and\
  \bibinfo {author} {\bibfnamefont {A.~M.}\ \bibnamefont {de~Paula}},\ }\href
  {\doibase 10.1103/PhysRevB.87.201401} {\bibfield  {journal} {\bibinfo
  {journal} {Phys. Rev. B}\ }\textbf {\bibinfo {volume} {87}},\ \bibinfo
  {pages} {201401} (\bibinfo {year} {2013})}\BibitemShut {NoStop}%
\bibitem [{\citenamefont {Janisch}\ \emph {et~al.}(2014)\citenamefont
  {Janisch}, \citenamefont {Wang}, \citenamefont {Ma}, \citenamefont {Mehta},
  \citenamefont {El{\'i}as}, \citenamefont {Perea-L{\'o}pez}, \citenamefont
  {Terrones}, \citenamefont {Crespi},\ and\ \citenamefont {Liu}}]{Janisch2014}%
  \BibitemOpen
  \bibfield  {author} {\bibinfo {author} {\bibfnamefont {C.}~\bibnamefont
  {Janisch}}, \bibinfo {author} {\bibfnamefont {Y.}~\bibnamefont {Wang}},
  \bibinfo {author} {\bibfnamefont {D.}~\bibnamefont {Ma}}, \bibinfo {author}
  {\bibfnamefont {N.}~\bibnamefont {Mehta}}, \bibinfo {author} {\bibfnamefont
  {A.~L.}\ \bibnamefont {El{\'i}as}}, \bibinfo {author} {\bibfnamefont
  {N.}~\bibnamefont {Perea-L{\'o}pez}}, \bibinfo {author} {\bibfnamefont
  {M.}~\bibnamefont {Terrones}}, \bibinfo {author} {\bibfnamefont
  {V.}~\bibnamefont {Crespi}}, \ and\ \bibinfo {author} {\bibfnamefont
  {Z.}~\bibnamefont {Liu}},\ }\href {http://dx.doi.org/10.1038/srep05530}
  {\bibfield  {journal} {\bibinfo  {journal} {Sci. Rep.}\ }\textbf
  {\bibinfo {volume} {4}},\ \bibinfo {pages} {5530} (\bibinfo {year}
  {2014})}\BibitemShut {NoStop}%
\bibitem [{\citenamefont {Rivera}\ \emph {et~al.}(2015)\citenamefont {Rivera},
  \citenamefont {Schaibley}, \citenamefont {Jones}, \citenamefont {Ross},
  \citenamefont {Wu}, \citenamefont {Aivazian}, \citenamefont {Klement},
  \citenamefont {Seyler}, \citenamefont {Clark}, \citenamefont {Ghimire},
  \citenamefont {Yan}, \citenamefont {Mandrus}, \citenamefont {Yao},\ and\
  \citenamefont {Xu}}]{Rivera2015}%
  \BibitemOpen
  \bibfield  {author} {\bibinfo {author} {\bibfnamefont {P.}~\bibnamefont
  {Rivera}}, \bibinfo {author} {\bibfnamefont {J.~R.}\ \bibnamefont
  {Schaibley}}, \bibinfo {author} {\bibfnamefont {A.~M.}\ \bibnamefont
  {Jones}}, \bibinfo {author} {\bibfnamefont {J.~S.}\ \bibnamefont {Ross}},
  \bibinfo {author} {\bibfnamefont {S.}~\bibnamefont {Wu}}, \bibinfo {author}
  {\bibfnamefont {G.}~\bibnamefont {Aivazian}}, \bibinfo {author}
  {\bibfnamefont {P.}~\bibnamefont {Klement}}, \bibinfo {author} {\bibfnamefont
  {K.}~\bibnamefont {Seyler}}, \bibinfo {author} {\bibfnamefont
  {G.}~\bibnamefont {Clark}}, \bibinfo {author} {\bibfnamefont {N.~J.}\
  \bibnamefont {Ghimire}}, \bibinfo {author} {\bibfnamefont {J.}~\bibnamefont
  {Yan}}, \bibinfo {author} {\bibfnamefont {D.~G.}\ \bibnamefont {Mandrus}},
  \bibinfo {author} {\bibfnamefont {W.}~\bibnamefont {Yao}}, \ and\ \bibinfo
  {author} {\bibfnamefont {X.}~\bibnamefont {Xu}},\ }\href
  {http://dx.doi.org/10.1038/ncomms7242} {\bibfield  {journal} {\bibinfo
  {journal} {Nat. Commun.}\ }\textbf {\bibinfo {volume} {6}},\
  \bibinfo {pages} {6242} (\bibinfo {year} {2015})}\BibitemShut {NoStop}%
\bibitem [{\citenamefont {Nayak}\ \emph {et~al.}(2017)\citenamefont {Nayak},
  \citenamefont {Horbatenko}, \citenamefont {Ahn}, \citenamefont {Kim},
  \citenamefont {Lee}, \citenamefont {Ma}, \citenamefont {Jang}, \citenamefont
  {Lim}, \citenamefont {Kim}, \citenamefont {Ryu}, \citenamefont {Cheong},
  \citenamefont {Park},\ and\ \citenamefont {Shin}}]{Nayak2017}%
  \BibitemOpen
  \bibfield  {author} {\bibinfo {author} {\bibfnamefont {P.~K.}\ \bibnamefont
  {Nayak}}, \bibinfo {author} {\bibfnamefont {Y.}~\bibnamefont {Horbatenko}},
  \bibinfo {author} {\bibfnamefont {S.}~\bibnamefont {Ahn}}, \bibinfo {author}
  {\bibfnamefont {G.}~\bibnamefont {Kim}}, \bibinfo {author} {\bibfnamefont
  {J.-U.}\ \bibnamefont {Lee}}, \bibinfo {author} {\bibfnamefont {K.~Y.}\
  \bibnamefont {Ma}}, \bibinfo {author} {\bibfnamefont {A.-R.}\ \bibnamefont
  {Jang}}, \bibinfo {author} {\bibfnamefont {H.}~\bibnamefont {Lim}}, \bibinfo
  {author} {\bibfnamefont {D.}~\bibnamefont {Kim}}, \bibinfo {author}
  {\bibfnamefont {S.}~\bibnamefont {Ryu}}, \bibinfo {author} {\bibfnamefont
  {H.}~\bibnamefont {Cheong}}, \bibinfo {author} {\bibfnamefont
  {N.}~\bibnamefont {Park}}, \ and\ \bibinfo {author} {\bibfnamefont {H.~S.}\
  \bibnamefont {Shin}},\ }\href {https://doi.org/10.1021/acsnano.7b00640}
  {\bibfield  {journal} {\bibinfo  {journal} {ACS Nano}\ }\textbf {\bibinfo
  {volume} {11}},\ \bibinfo {pages} {4041} (\bibinfo {year}
  {2017})}\BibitemShut {NoStop}%
\bibitem [{\citenamefont {Mudd}\ \emph {et~al.}(2016)\citenamefont {Mudd},
  \citenamefont {Molas}, \citenamefont {Chen}, \citenamefont {Z{\'o}lyomi},
  \citenamefont {Nogajewski}, \citenamefont {Kudrynskyi}, \citenamefont
  {Kovalyuk}, \citenamefont {Yusa}, \citenamefont {Makarovsky}, \citenamefont
  {Eaves}, \citenamefont {Potemski}, \citenamefont {Fal'ko},\ and\
  \citenamefont {Patan{\`e}}}]{Mudd2016}%
  \BibitemOpen
  \bibfield  {author} {\bibinfo {author} {\bibfnamefont {G.~W.}\ \bibnamefont
  {Mudd}}, \bibinfo {author} {\bibfnamefont {M.~R.}\ \bibnamefont {Molas}},
  \bibinfo {author} {\bibfnamefont {X.}~\bibnamefont {Chen}}, \bibinfo {author}
  {\bibfnamefont {V.}~\bibnamefont {Z{\'o}lyomi}}, \bibinfo {author}
  {\bibfnamefont {K.}~\bibnamefont {Nogajewski}}, \bibinfo {author}
  {\bibfnamefont {Z.~R.}\ \bibnamefont {Kudrynskyi}}, \bibinfo {author}
  {\bibfnamefont {Z.~D.}\ \bibnamefont {Kovalyuk}}, \bibinfo {author}
  {\bibfnamefont {G.}~\bibnamefont {Yusa}}, \bibinfo {author} {\bibfnamefont
  {O.}~\bibnamefont {Makarovsky}}, \bibinfo {author} {\bibfnamefont
  {L.}~\bibnamefont {Eaves}}, \bibinfo {author} {\bibfnamefont
  {M.}~\bibnamefont {Potemski}}, \bibinfo {author} {\bibfnamefont {V.~I.}\
  \bibnamefont {Fal'ko}}, \ and\ \bibinfo {author} {\bibfnamefont
  {A.}~\bibnamefont {Patan{\`e}}},\ }\href
  {http://dx.doi.org/10.1038/srep39619} {\bibfield  {journal} {\bibinfo
  {journal} {Sci. Rep.}\ }\textbf {\bibinfo {volume} {6}},\ \bibinfo
  {pages} {39619} (\bibinfo {year} {2016})}\BibitemShut {NoStop}%
\bibitem [{\citenamefont {Bandurin}\ \emph {et~al.}(2016)\citenamefont
  {Bandurin}, \citenamefont {Tyurnina}, \citenamefont {Yu}, \citenamefont
  {Mishchenko}, \citenamefont {Z{\'o}lyomi}, \citenamefont {Morozov},
  \citenamefont {Kumar}, \citenamefont {Gorbachev}, \citenamefont {Kudrynskyi},
  \citenamefont {Pezzini}, \citenamefont {Kovalyuk}, \citenamefont {Zeitler},
  \citenamefont {Novoselov}, \citenamefont {Patan{\`e}}, \citenamefont {Eaves},
  \citenamefont {Grigorieva}, \citenamefont {Fal'ko}, \citenamefont {Geim},\
  and\ \citenamefont {Cao}}]{Bandurin2016}%
  \BibitemOpen
  \bibfield  {author} {\bibinfo {author} {\bibfnamefont {D.~A.}\ \bibnamefont
  {Bandurin}}, \bibinfo {author} {\bibfnamefont {A.~V.}\ \bibnamefont
  {Tyurnina}}, \bibinfo {author} {\bibfnamefont {G.~L.}\ \bibnamefont {Yu}},
  \bibinfo {author} {\bibfnamefont {A.}~\bibnamefont {Mishchenko}}, \bibinfo
  {author} {\bibfnamefont {V.}~\bibnamefont {Z{\'o}lyomi}}, \bibinfo {author}
  {\bibfnamefont {S.~V.}\ \bibnamefont {Morozov}}, \bibinfo {author}
  {\bibfnamefont {R.~K.}\ \bibnamefont {Kumar}}, \bibinfo {author}
  {\bibfnamefont {R.~V.}\ \bibnamefont {Gorbachev}}, \bibinfo {author}
  {\bibfnamefont {Z.~R.}\ \bibnamefont {Kudrynskyi}}, \bibinfo {author}
  {\bibfnamefont {S.}~\bibnamefont {Pezzini}}, \bibinfo {author} {\bibfnamefont
  {Z.~D.}\ \bibnamefont {Kovalyuk}}, \bibinfo {author} {\bibfnamefont
  {U.}~\bibnamefont {Zeitler}}, \bibinfo {author} {\bibfnamefont {K.~S.}\
  \bibnamefont {Novoselov}}, \bibinfo {author} {\bibfnamefont {A.}~\bibnamefont
  {Patan{\`e}}}, \bibinfo {author} {\bibfnamefont {L.}~\bibnamefont {Eaves}},
  \bibinfo {author} {\bibfnamefont {I.~V.}\ \bibnamefont {Grigorieva}},
  \bibinfo {author} {\bibfnamefont {V.~I.}\ \bibnamefont {Fal'ko}}, \bibinfo
  {author} {\bibfnamefont {A.~K.}\ \bibnamefont {Geim}}, \ and\ \bibinfo
  {author} {\bibfnamefont {Y.}~\bibnamefont {Cao}},\ }\href
  {http://dx.doi.org/10.1038/nnano.2016.242} {\bibfield  {journal} {\bibinfo
  {journal} {Nat. Nanotechnol.}\ }\textbf {\bibinfo {volume} {12}},\
  \bibinfo {pages} {223} (\bibinfo {year} {2016})}\BibitemShut {NoStop}%
\bibitem [{\citenamefont {Deckoff-Jones}\ \emph {et~al.}(2016)\citenamefont
  {Deckoff-Jones}, \citenamefont {Zhang}, \citenamefont {Petoukhoff},
  \citenamefont {Man}, \citenamefont {Lei}, \citenamefont {Vajtai},
  \citenamefont {Ajayan}, \citenamefont {Talbayev}, \citenamefont {Mad{\'e}o},\
  and\ \citenamefont {Dani}}]{Deckoff2016}%
  \BibitemOpen
  \bibfield  {author} {\bibinfo {author} {\bibfnamefont {S.}~\bibnamefont
  {Deckoff-Jones}}, \bibinfo {author} {\bibfnamefont {J.}~\bibnamefont
  {Zhang}}, \bibinfo {author} {\bibfnamefont {C.~E.}\ \bibnamefont
  {Petoukhoff}}, \bibinfo {author} {\bibfnamefont {M.~K.~L.}\ \bibnamefont
  {Man}}, \bibinfo {author} {\bibfnamefont {S.}~\bibnamefont {Lei}}, \bibinfo
  {author} {\bibfnamefont {R.}~\bibnamefont {Vajtai}}, \bibinfo {author}
  {\bibfnamefont {P.~M.}\ \bibnamefont {Ajayan}}, \bibinfo {author}
  {\bibfnamefont {D.}~\bibnamefont {Talbayev}}, \bibinfo {author}
  {\bibfnamefont {J.}~\bibnamefont {Mad{\'e}o}}, \ and\ \bibinfo {author}
  {\bibfnamefont {K.~M.}\ \bibnamefont {Dani}},\ }\href
  {http://dx.doi.org/10.1038/srep22620} {\bibfield  {journal} {\bibinfo
  {journal} {Sci. Rep.}\ }\textbf {\bibinfo {volume} {6}},\ \bibinfo
  {pages} {22620} (\bibinfo {year} {2016})}\BibitemShut {NoStop}%
\bibitem [{\citenamefont {Pozo-Zamudio}\ \emph {et~al.}(2015)\citenamefont
  {Pozo-Zamudio}, \citenamefont {Schwarz}, \citenamefont {Klein}, \citenamefont
  {Schofield}, \citenamefont {Chekhovich}, \citenamefont {Ceylan},
  \citenamefont {Margapoti}, \citenamefont {Dmitriev}, \citenamefont
  {Lashkarev}, \citenamefont {Borisenko}, \citenamefont {Kolesnikov},
  \citenamefont {Finley},\ and\ \citenamefont {Tartakovskii}}]{Zamudio2015}%
  \BibitemOpen
  \bibfield  {author} {\bibinfo {author} {\bibfnamefont {O.~D.}\ \bibnamefont
  {Pozo-Zamudio}}, \bibinfo {author} {\bibfnamefont {S.}~\bibnamefont
  {Schwarz}}, \bibinfo {author} {\bibfnamefont {J.}~\bibnamefont {Klein}},
  \bibinfo {author} {\bibfnamefont {R.~C.}\ \bibnamefont {Schofield}}, \bibinfo
  {author} {\bibfnamefont {E.~A.}\ \bibnamefont {Chekhovich}}, \bibinfo
  {author} {\bibfnamefont {O.}~\bibnamefont {Ceylan}}, \bibinfo {author}
  {\bibfnamefont {E.}~\bibnamefont {Margapoti}}, \bibinfo {author}
  {\bibfnamefont {A.~I.}\ \bibnamefont {Dmitriev}}, \bibinfo {author}
  {\bibfnamefont {G.~V.}\ \bibnamefont {Lashkarev}}, \bibinfo {author}
  {\bibfnamefont {D.~N.}\ \bibnamefont {Borisenko}}, \bibinfo {author}
  {\bibfnamefont {N.~N.}\ \bibnamefont {Kolesnikov}}, \bibinfo {author}
  {\bibfnamefont {J.~J.}\ \bibnamefont {Finley}}, \ and\ \bibinfo {author}
  {\bibfnamefont {A.~I.}\ \bibnamefont {Tartakovskii}},\ }\href@noop {}
  {\bibfield  {journal} {\bibinfo  {journal} {arXiv:1506.05619}\ } (\bibinfo
  {year} {2015})}\BibitemShut {NoStop}%
\bibitem [{\citenamefont {Cao}\ \emph {et~al.}(2015)\citenamefont {Cao},
  \citenamefont {Mishchenko}, \citenamefont {Yu}, \citenamefont {Khestanova},
  \citenamefont {Rooney}, \citenamefont {Prestat}, \citenamefont {Kretinin},
  \citenamefont {Blake}, \citenamefont {Shalom}, \citenamefont {Woods},
  \citenamefont {Chapman}, \citenamefont {Balakrishnan}, \citenamefont
  {Grigorieva}, \citenamefont {Novoselov}, \citenamefont {Piot}, \citenamefont
  {Potemski}, \citenamefont {Watanabe}, \citenamefont {Taniguchi},
  \citenamefont {Haigh}, \citenamefont {Geim},\ and\ \citenamefont
  {Gorbachev}}]{Cao2015}%
  \BibitemOpen
  \bibfield  {author} {\bibinfo {author} {\bibfnamefont {Y.}~\bibnamefont
  {Cao}}, \bibinfo {author} {\bibfnamefont {A.}~\bibnamefont {Mishchenko}},
  \bibinfo {author} {\bibfnamefont {G.~L.}\ \bibnamefont {Yu}}, \bibinfo
  {author} {\bibfnamefont {E.}~\bibnamefont {Khestanova}}, \bibinfo {author}
  {\bibfnamefont {A.~P.}\ \bibnamefont {Rooney}}, \bibinfo {author}
  {\bibfnamefont {E.}~\bibnamefont {Prestat}}, \bibinfo {author} {\bibfnamefont
  {A.~V.}\ \bibnamefont {Kretinin}}, \bibinfo {author} {\bibfnamefont
  {P.}~\bibnamefont {Blake}}, \bibinfo {author} {\bibfnamefont {M.~B.}\
  \bibnamefont {Shalom}}, \bibinfo {author} {\bibfnamefont {C.}~\bibnamefont
  {Woods}}, \bibinfo {author} {\bibfnamefont {J.}~\bibnamefont {Chapman}},
  \bibinfo {author} {\bibfnamefont {G.}~\bibnamefont {Balakrishnan}}, \bibinfo
  {author} {\bibfnamefont {I.~V.}\ \bibnamefont {Grigorieva}}, \bibinfo
  {author} {\bibfnamefont {K.~S.}\ \bibnamefont {Novoselov}}, \bibinfo {author}
  {\bibfnamefont {B.~A.}\ \bibnamefont {Piot}}, \bibinfo {author}
  {\bibfnamefont {M.}~\bibnamefont {Potemski}}, \bibinfo {author}
  {\bibfnamefont {K.}~\bibnamefont {Watanabe}}, \bibinfo {author}
  {\bibfnamefont {T.}~\bibnamefont {Taniguchi}}, \bibinfo {author}
  {\bibfnamefont {S.~J.}\ \bibnamefont {Haigh}}, \bibinfo {author}
  {\bibfnamefont {A.~K.}\ \bibnamefont {Geim}}, \ and\ \bibinfo {author}
  {\bibfnamefont {R.~V.}\ \bibnamefont {Gorbachev}},\ }\href {\doibase
  10.1021/acs.nanolett.5b00648} {\bibfield  {journal} {\bibinfo  {journal}
  {Nano Lett.}\ }\textbf {\bibinfo {volume} {15}},\ \bibinfo {pages} {4914}
  (\bibinfo {year} {2015})}\BibitemShut {NoStop}%
\bibitem [{\citenamefont {Yariv}\ and\ \citenamefont {Yeh}(1984)}]{Yariv1984}%
  \BibitemOpen
  \bibfield  {author} {\bibinfo {author} {\bibfnamefont {A.}~\bibnamefont
  {Yariv}}\ and\ \bibinfo {author} {\bibfnamefont {P.}~\bibnamefont {Yeh}},\
  }\enquote {\bibinfo {title} {Optical waves in crystals: propagation and
  control of laser radiation},}\ \ (\bibinfo  {publisher} {Wiley},\ \bibinfo
  {address} {New York},\ \bibinfo {year} {1984})\ Chap.~\bibinfo {chapter}
  {12}\BibitemShut {NoStop}%
\bibitem [{\citenamefont {Shen}(1989)}]{Shen1989}%
  \BibitemOpen
  \bibfield  {author} {\bibinfo {author} {\bibfnamefont {Y.~R.}\ \bibnamefont
  {Shen}},\ }\href {\doibase 10.1146/annurev.pc.40.100189.001551} {\bibfield
  {journal} {\bibinfo  {journal} {Annu. Rev. Phys. Chem.}\
  }\textbf {\bibinfo {volume} {40}},\ \bibinfo {pages} {327} (\bibinfo {year}
  {1989})}\BibitemShut {NoStop}%
\bibitem [{\citenamefont {Tang}\ \emph {et~al.}(2016)\citenamefont {Tang},
  \citenamefont {Mandal}, \citenamefont {McGuire},\ and\ \citenamefont
  {Lai}}]{Tang2016}%
  \BibitemOpen
  \bibfield  {author} {\bibinfo {author} {\bibfnamefont {Y.}~\bibnamefont
  {Tang}}, \bibinfo {author} {\bibfnamefont {K.~C.}\ \bibnamefont {Mandal}},
  \bibinfo {author} {\bibfnamefont {J.~A.}\ \bibnamefont {McGuire}}, \ and\
  \bibinfo {author} {\bibfnamefont {C.~W.}\ \bibnamefont {Lai}},\ }\href
  {https://link.aps.org/doi/10.1103/PhysRevB.94.125302} {\bibfield  {journal}
  {\bibinfo  {journal} {Phys. Rev. B}\ }\textbf {\bibinfo {volume} {94}},\
  \bibinfo {pages} {125302} (\bibinfo {year} {2016})}\BibitemShut {NoStop}%
\end{thebibliography}

\end{document}